\documentclass[preprint]{aastex} 
\usepackage{color}
\usepackage{graphicx}
\usepackage{epsfig}
\usepackage{natbib}
\usepackage{rotating}
\usepackage{amsmath}

\begin{document}

\title{Demonstration of a Near-IR Laser Comb for Precision Radial Velocity Measurements in Astronomy}

\author{X. Yi$^1$,K. Vahala*$^1$,J. Li$^1$,
S. Diddams$^{2,3}$,G. Ycas$^{2,3}$,
P. Plavchan$^4$,
S. Leifer$^5$,J. Sandhu$^5$,G. Vasisht$^5$,P. Chen$^5$,
P. Gao$^6$,
J. Gagne$^7$,
E. Furlan$^8$,M. Bottom$^8$,
E. Martin$^9$,M. Fitzgerald$^9$,
G. Doppmann$^{10}$,
C. Beichman*$^{10}$}

\affil{1) Laboratory of Applied Physics,California Institute of Technology, Pasadena, CA 91125}
\affil{2) National Institute of Standards and Technology, 325 Broadway, Boulder, CO 80305}
\affil{3) Department of Physics, University of Colorado, 2000 Colorado Ave, Boulder, CO, 80309}
\affil{4) Dept. of Physics, Missouri State University, 901 S National Ave, Springfield, MO 65897}
\affil{5) Jet Propulsion Laboratory, 4800 Oak Grove Drive, Pasadena, CA 91125}
\affil{6) Division of Geological and Planetary Sciences, California Institute of Technology, Pasadena, CA 91125}
\affil{7) Department of Physics, University of  Montreal, Montreal, Canada}
\affil{8) Department of Astronomy, California Institute of Technology, Pasadena, CA 91125}
\affil{9) Department of Physics and Astronomy, University of  California Los Angeles, Los Angeles, CA 90095}
\affil{10) W.M. Keck Observatory, Kamuela, HI 96743}
\affil{11) NASA Exoplanet Science Institute, California Institute of Technology, Pasadena, CA 91125}
\affil{*) Corresponding authors: chas *at* ipac.caltech.edu, vahala *at* caltech.edu }

\begin{abstract}
We describe a successful effort to produce a laser comb around 1.55 $\mu$m in the astronomical H band (1.5-1.8 $\mu$m) using a method based on a line-referenced, electro-optical-modulation frequency comb (LR-EOFC).  We discuss the experimental setup, laboratory results and proof of concept demonstrations at the NASA Infrared Telescope Facility (IRTF) and the Keck-II telescope. The laser comb has a demonstrated stability of $<$ 200 kHz, corresponding to a Doppler precision of $\sim 0.3$ m s$^{-1}$. This technology, when coupled with a high spectral resolution spectrograph, offers the promise of $\sim$ 1 m s$^{-1}$ radial velocity precision suitable for the detection of Earth-sized planets in the habitable zones of cool M-type stars.
\end{abstract}

\section{Introduction}

The earliest technique for the discovery and characterization of planets orbiting other stars (``exoplanets") is the Doppler or Radial Velocity (RV) method whereby small periodic changes in the motion of a star orbited by a planet are detected via careful spectroscopic measurements \citep{Perryman2011}. The RV technique has identified hundreds of planets ranging in mass from a few times the mass of Jupiter to less than an Earth mass, and in orbital periods from less than a day to over 10 years \citep{Marcy2011}. However, the detection of Earth-analogs at orbital separations suitable for the presence of liquid water at the planet's surface, i.e. in the "Habitable Zone" (HZ) \citep{Kasting1993}, remains challenging for stars like the Sun with RV  signatures $<$ 0.1  m s$^{-1}$ ($\Delta V/c< 3\times10^{-10}$) and periods of a year ($\sim$ 10$^8$ sec to measure three complete periods). For cooler, lower luminosity stars (spectral class M), however, the Habitable Zone moves closer to the star which, by application of Kepler's laws, implies that a planet's radial velocity signature increases,  $\sim$ 0.5  m s$^{-1}$ ($\Delta V/c< 1.5\times10^{-9}$), and its  orbital period decreases, $\sim$ 30 days ($\sim10^7$ sec to measure three periods). Both of these effects make the detection problem easier. But for M stars the bulk of the radiation shifts from the visible wavelengths, where most RV measurements have been made to date, into the near-infrared. Thus, there is considerable interest among astronomers in developing precise RV capabilities at longer wavelengths.

Critical to precision RV measurements is a highly stable wavelength reference. Recently a number of groups have undertaken to provide an alternative calibration standard that consists of a ``comb" of evenly spaced laser lines accurately anchored to a stable frequency standard and injected directly into the spectrometer along with the stellar spectrum  \cite{Murphy2007, Osterman2007, Li2008, Braje2008}. While this effort has mostly been focused on  visible wavelengths, there have been successful efforts at near-IR wavelengths as well \citep{Steinmetz2008, Ycas2012a, Quinlan2010}.  In all of these earlier studies, the comb has been based on a femtosecond mode-locked laser that is self-referenced \citep{Jones2000}, such that  the spectral line spacing and common offset frequency of all lines are both locked to a radio frequency standard.  Thus, laser combs potentially represent an ideal tool for spectroscopic measurements and radial velocity measurement. 

However, in the case of mode-locked laser combs, the line spacing is typically in the range of 0.1-1 GHz, which is too small to be resolved by most astronomical spectrographs.  As a result, the output spectrum of the comb must be spectrally filtered to create a calibration grid spaced by $>$10 GHz, which is more commensurate with the resolving power of a high-resolution astronomical spectrograph \citep{Braje2008}. While this approach has led to calibration results at the cm/sec level \citep{Wilken2012}, it nonetheless increases the complexity and cost of the system.  

 In light of this, there is interest in developing photonic tools that possess many of the benefits of mode-locked laser combs, but that might be simpler, less expensive, and more amenable to ``hands-off'' operation at remote telescope sites.  Indeed, in many radial velocity measurements, other system-induced errors and uncertainties can limit the achievable precision, such that a frequency comb of lesser precision could still be equally valuable.   For example, one alternative technique recently reported is to use a series of spectroscopic minima induced in a broad continuum spectrum using a compact Fabry-Perot interferometer \citep{Wildi2010, Halverson2014}. While the technique must account for temperature-induced tuning of the interferometer, it has the advantage of simplicity and low cost.  Another interesting alternative is the so-called Kerr comb or microcomb, which have the distinct advantage of directly providing a comb with spacing in the range of 10-100 GHz, without the need for filtering \citep{Kippenberg2011}.  While this new type of laser comb is still under development, there have been promising demonstrations of full microcomb frequency control\citep{DelHaye2008,Papp2014} and in the future it could be possible to fully integrate such a microcomb on only a few square centimeters of silicon, making a very robust and inexpensive calibrator. 

Another approach that has been proposed is to create a comb through electro-optical modulation of a frequency stabilized laser \citep{Suziki2013}. In this paper, we describe a successful effort to implement this approach. We produce a laser comb around 1.55 $\mu$m in the astronomical H band (1.5-1.8 $\mu$m) based on a line-referenced, electro-optical-modulation frequency comb (LR-EOFC). We discuss the experimental setup, laboratory results and proof of concept demonstrations at the NASA Infrared Telescope Facility (IRTF) and the WMKO Keck 10 m telescope.

\section{Description of Line-Referenced, Electro-Optic Frequency Comb System (LR-EOFC)}

\subsection{Background of Electro-Optic Modulation for Comb Generation}
\begin{figure}[htbp]
\centering
\includegraphics[width=\linewidth,natwidth=4768,natheight=3160]{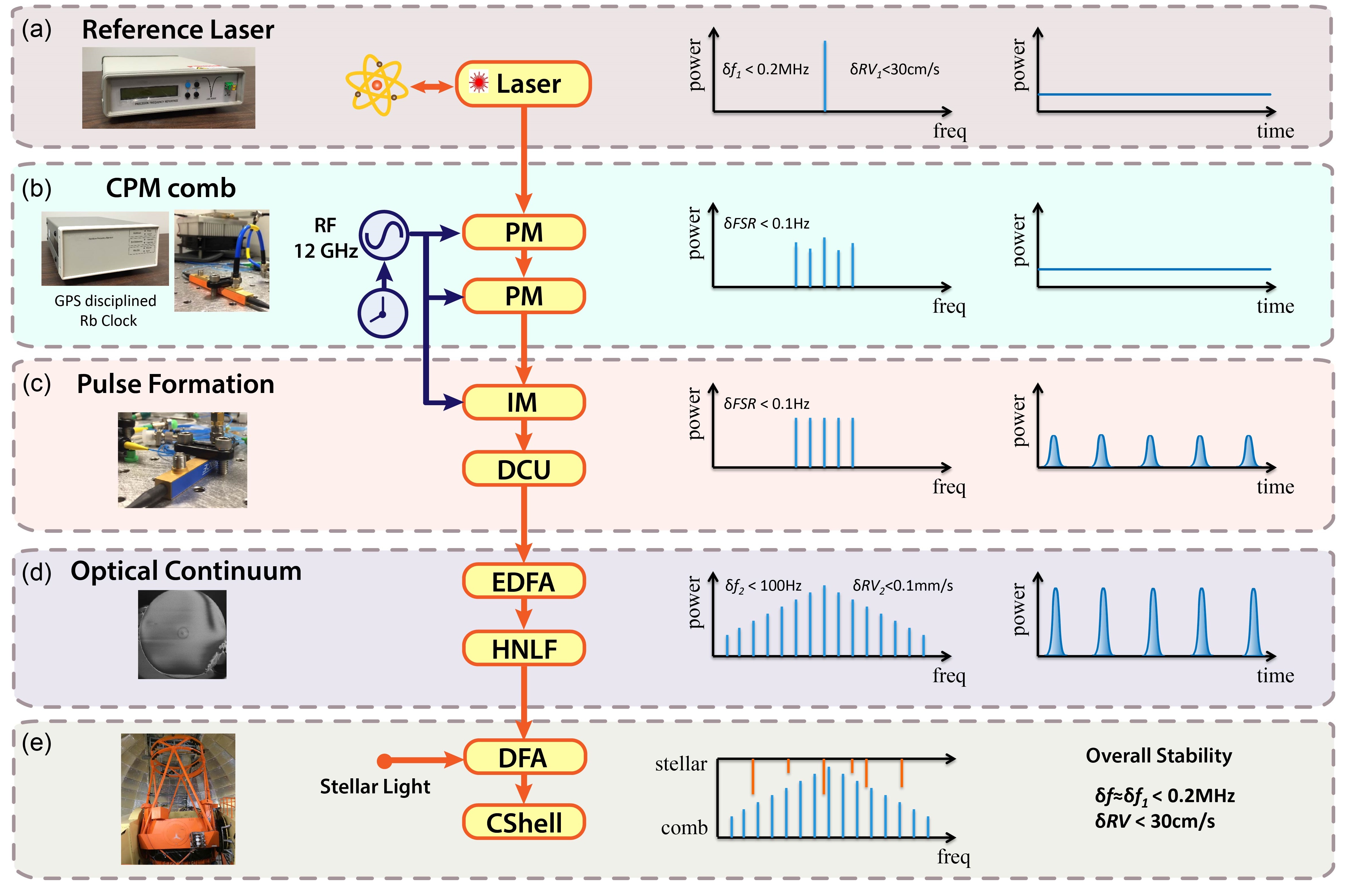}
\caption{Conceptual schematics of EOM astro-comb. Vertically, the first column contains photographs of key instruments. A simplified schematic setup is in second column. Third and forth columns present the comb state in the frequency and temporal domains. (a) The pump laser is locked to a molecular transition, acquiring stability of 0.2MHz, corresponding to 30cm/s RV. (b) The phase of this laser is then modulated by two tandem phase modulators (PM), creating several tens of sidebands with spacing equal to modulation frequency. (c) Pulse forming is then performed by an intensity modulator (IM) and dispersion compensation unit (DCU), which could be a piece of proper length single mode fiber (SMF) or a fiber Bragg grating. (d) After amplification by an erbium doped fiber amplifier (EDFA), the pulse undergoes optical continuum generation in a highly nonlinear fiber (HNLF), extending its bandwidth $>100$nm. (e) The comb is combined with stellar light by using two fiber arrays and is sent into telescope spectrograph. The overall comb stability is determined by the pump laser.}
\label{figure2}
\end{figure}

An Electro-Optic Frequency Comb System (EOFC) \citep{Murata2000, Metcalf2013}  is generated in a distinctly different way compared to a mode-locked laser comb. The procedure with depictions of corresponding spectral and time-domain signals is illustrated in Fig. \ref{figure2}.  The comb center frequency is set by a frequency stabilized pump laser, $\nu_o$, which is then phase modulated at a microwave frequency, $f_m$, to produce phase modulation sidebands $\nu =\nu_o \pm nf_m$, where $n$ is an integer.  The sidebands have a spacing equal to the modulation frequency and are therefore tunable and easily set to provide a spectral line spacing that can be resolved by an astronomical spectrometer.

Recently, the EOFC approach has been applied to pulse generation and subsequent continuum generation in nonlinear optical fibers, yielding combs with spacing in the range of 10-25 GHz that span hundreds of nanometers around 1550 nm \citep{Ishizawa2011, Kourogi1998}. A line-referenced EOFC (LR-EOFC) is simply a spectrum of lines generated by electro-optic modulation, where the frequency of the CW laser source has been stabilized to an atomic or molecular reference (e.g. $\nu_o = \nu_{atom}$). In this case, the uncertainty in the position of the comb teeth is determined by the combination of the uncertainty in the stabilization of $\nu_o$ and the uncertainty of the microwave source that provides the modulation frequency $f_m$.  However, the typical uncertainty of a microwave source can be sub-Hertz when synchronized with a compact Rb clock and moreover can be GPS-disciplined to provide long-term stability \citep{Quinlan2010}.  Thus, the dominant uncertainty in comb tooth frequency in the LR-EOFC is that of $\nu_o$. Here, we have found that commercial frequency stabilized laser sources (gas cell stabilization) can be stable at the level of $\sim$200 kHz, which corresponds to an RV resolution as low as 30 cm s$^{-1}$ in H-band. While the LR-EOFC may not provide the same level of accuracy and precision that is possible from a self-referenced, mode-locked laser comb, it is nonetheless relatively simple to build, has a line spacing that is commensurate with astronomical spectrometers, and employs commercial-off-the-shelf (COTS) technologies developed for the 1550 nm telecommunications industry. The last fact makes the device relatively inexpensive in comparison to a conventional frequency comb and endows the device with high reliability - a critical factor in operation at remote telescope sites. Finally, the technique generalizes to other wavelength bands where optical amplifiers and phase modulators are available. For example, LR-EOFC comb generation should also be possible near 1 $\mu$m.

\subsection{Description of The Comb System}

The system studied in this work could operate around two center wavelengths provided by two commercially-available, frequency-stabilized lasers.  The entire LR-EOFC system was mounted on an aluminum breadboard (18" $\times$ 32") for transport and implementation with the CSHELL spectrograph at the NASA Infrared telescope facility (IRTF) on Mauna Kea in Hawaii \citep{Greene1993}. In addition, a preliminary coupling test of the comb system to the NIRSPEC spectrograph at Keck II in Hawaii was performed (i.e., no observations). Optical hardware was fiber pigtailed and polarization maintaining optical fiber was used for added stability.  On the breadboard, the frequency-stabilized laser was first coupled into two, tandem phase modulators with $V_\pi \sim 3.7$V. The phase modulators were driven by an amplified 12 GHz frequency generator and synchronized by using microwave phase shifters. This initial phase modulation process produced a comb having approximately a 4 nm bandwidth.  This initial comb was then coupled into an intensity modulator driven at the same microwave frequency that was applied to the phase modulators. Microwave phase shifters were used to align the drive phase so that the intensity modulator gated-out only those portions of the phase modulation that were approximately linearly chirped with one sign (i.e., parabolic phase variation in time). A nearly transform-limited pulse was then formed when this parabolic phase variation was nullified by dispersion compensation using a fiber grating. A 2 ps FWHM pulse was measured after the fiber grating using an autocorrelator in the temporal domain. Owing to this pulse formation, the duty cycle of the pulse train reached above 40, boosting the peak intensity of the pulses.

\begin{figure}[htbp]
\centering
\includegraphics[scale=0.25,natwidth=1580,natheight=2018]{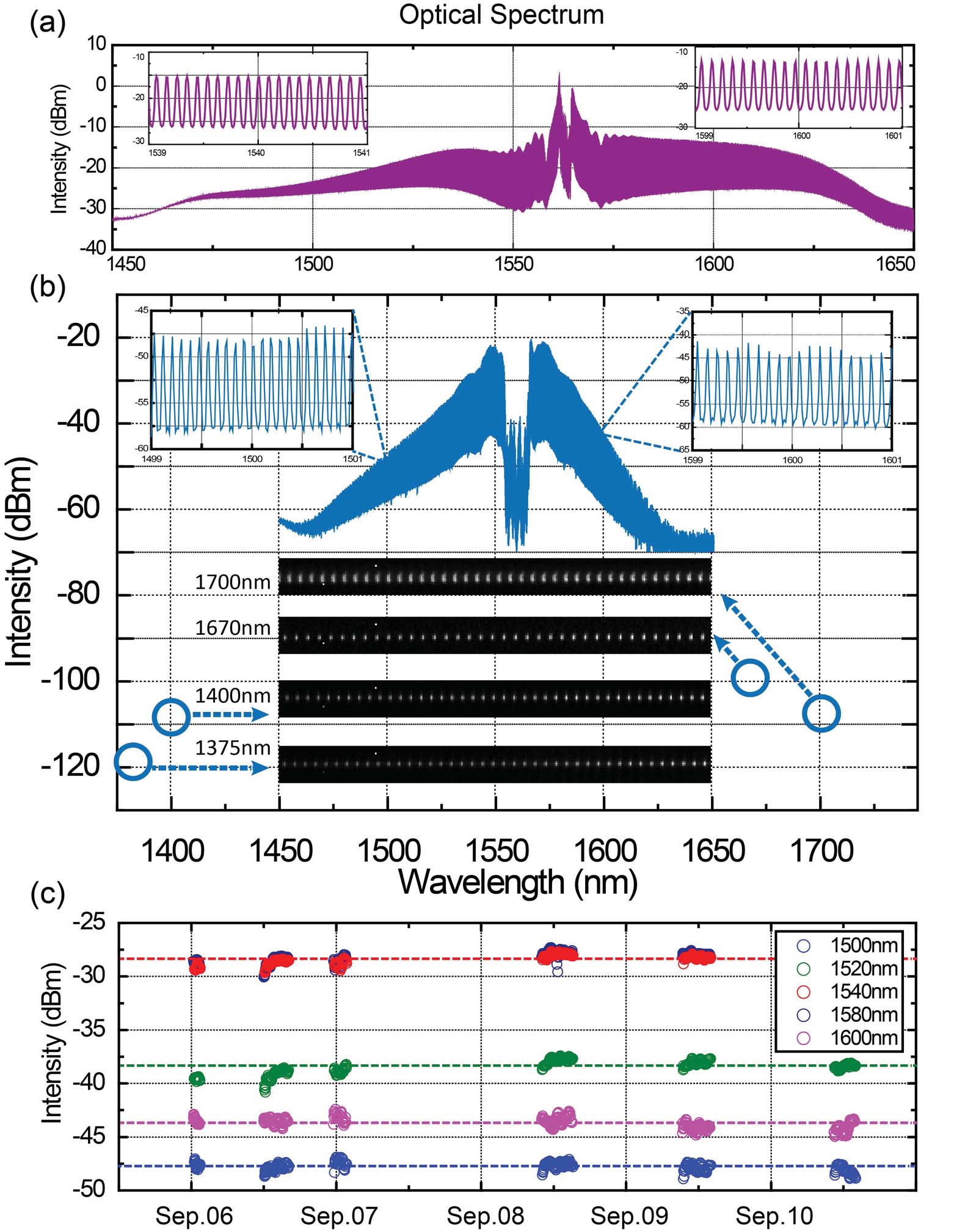}
\caption{Comb spectra. (a) A $>$100nm comb spectrum generated with 600mW pump power. The insets show the resolved line spacing of 12 GHz or approximately 0.1 nm. (b) Comb spectra generated with 150mW pump power. The insets on top left and right show the resolved comb lines. Comb spectra taken by the CSHELL spectrograph at 1375 nm, 1400 nm, 1670 nm and 1700 nm are presented as insets in the lower half of the figure. The blue circles mark the estimated comb line power and center wavelength for these spectra. Comb lines are detectable on CSHELL at fW power levels. (c) Comb spectral line power vs. time is shown at five different wavelengths. During the 5 day test at IRTF, no parameter adjustment was made, and comb intensity was very stable even with multiple power-on and off cycling of the optical continuum generation system.}
\label{figure4a}
\end{figure}

These pulses were then amplified in an Erbium Doped Fiber Amplifier (EDFA) and coupled into a 20 m length of highly-nonlinear fiber (HNLF). Propagation in the HNLF caused self-phase modulation and strong spectral broadening of the comb.  Because the CSHELL spectrometer at the IRTF facility has a spectral window of only 4 nm, there was no effort made to generate spectrally flat combs in this work. However, electro-optical-based combs that are wide band and have relatively flat spectra have been demonstrated \cite{Mori2005}. In this work, we generated comb spectra under several pumping conditions. A $>$100nm bandwidth comb with $>$10 $\mu$W power per tooth is recorded in fig. \ref{figure4a}a and was generated by pumping at a power level of 600 mW. Using the CSHELL spectrograph, we also measured a comb extending from 1375 nm to 1700 nm with pump power level of 150mW. A portion of the comb spectrum visible on a commercial spectrometer is shown in fig. \ref{figure4a}b. Narrow-band spectra taken using the highly sensitive CSHELL spectrograph are presented as insets in fig. \ref{figure4a}b with their spectral location and power levels indicated. This comb was generated to show that high contrast comb lines are visible over a very broad range of wavelengths using the EOFC  method. However, the dynamic range of power in these lines makes them unsuitable for spectral calibration. Nonetheless, by using different nonlinear fiber \citep{Mori2005} and some spectral flattening methods, broad combs with level power should be possible.

\subsection{Overall Stability of the LR-EOFC System}

\begin{figure}[htbp]
\centering
\includegraphics[width=\linewidth,natwidth=1683,natheight=1258]{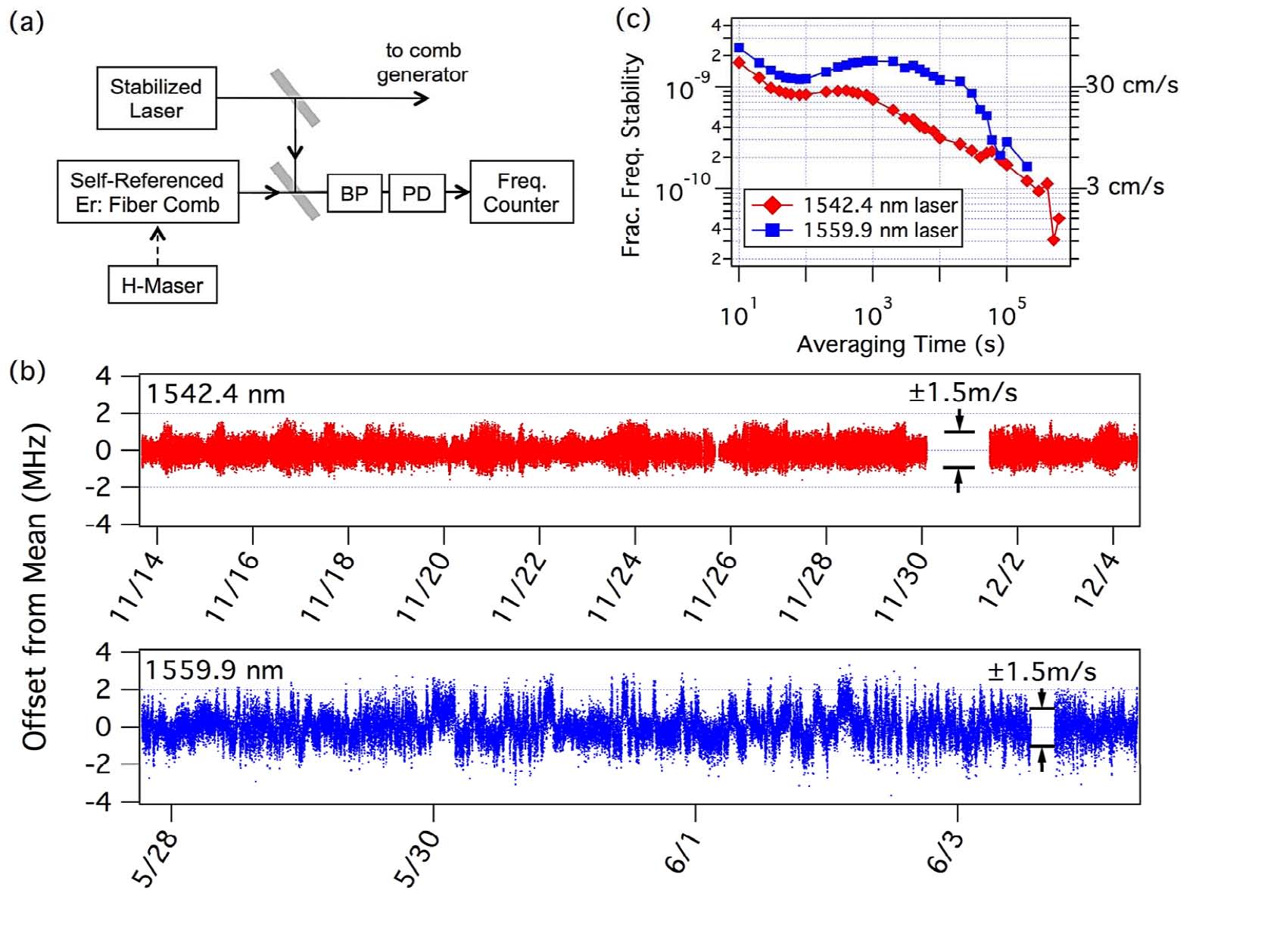}
\caption{Instability of the C$_2$H$_2$ and HCN reference lasers.  (a) Experimental setup.  BP, optical bandpass filter; PD, photodiode. All beam paths and beam combiners are in single mode fiber.  (b) Time series of measured beat frequencies for the two frequency-stabilized lasers with 10 s averaging per measurement. The x-axes are the dates in November of 2013 and May/June of 2014, respectively .  (c) Allan deviation, which is a measure of the fractional frequency instability, computed from the time series data of (b). Inset scales give the radial velocity precision.}
\label{AllanDev}
\end{figure}

As noted above, the frequency stability of the LR-EOFC is dominated by the stability of the reference laser $\nu_o$.  We explored the use of two different commercially-available lasers (Wavelength References) that were stabilized respectively to an acetylene (C$_2$H$_2$) line at 1542.4 nm, and to a hydrogen cyanide (H$^{13}$C$^{15}$N) line at 1559.9 nm. To assess the stability of the sources, their frequencies were measured relative to an Er:fiber based self-referenced optical frequency comb \citep{Ycas2012a, Ycas2012b}.  Fiber-coupled light from a reference laser was combined into a common optical fiber with light from the Er:fiber comb. Then the heterodyne beat between a single comb line and the line-stabilized reference was filtered, amplified, and counted with a 10 s gate time using a frequency counter that was synchronized to a hydrogen maser (see Fig \ref{AllanDev}(a)).  The Er:fiber comb was stabilized relative to the same hydrogen maser, such that the fractional frequency stability of the measurement was $<2\times10^{-13}$ at all averaging times.  The drift of the hydrogen maser frequency is $<1\times10^{-15}$ per day, thereby providing a stable reference at levels corresponding to a radial velocity uncertainty $\ll\,1$ cm/sec.  Thus, the frequency of the counted heterodyne beat accurately represents the fluctuations in the reference laser.  

The series of 10 s measurements of the heterodyne beat was recorded over 20 days in 2013 for the case of the 1542 nm laser and more than 7 days in 2014 for the case of the 1560 nm laser, as shown in Fig. \ref{AllanDev}(b).  Gaps in the measurements near 11/31 and 6/4 are due to unlocking of the Er:fiber comb from the hydrogen maser reference. From these time series, we calculate the Allan deviation, which is a measure of the fractional frequency fluctuations (instability) of the reference laser as a function of averaging time.  As seen in Fig. \ref{AllanDev}(c), the instability of the 1542 nm laser is $<\,10^{-9}$ (30 cm s$^{-1}$ RV) at all averaging times greater than $\sim\,$30 s.  The 1560 nm laser is less stable, but provides a corresponding RV precision of $<\,60$ cm s$^{-1}$ for averaging times greater than 20 s.  This difference in stability was to be expected because of the difference in relative absorption line strength between the acetylene and HCN stabilized lasers. In both cases, the stability improves with averaging time, although at a rate slower than predicted for white frequency noise.  We believe this to be the result of the modulation scheme employed to lock the reference laser to the relatively broad  center of the Doppler and pressure-broadened acetylene and HCN lines. 

Additional analysis included the measurement of the drift of the frequencies of the two reference lasers by fitting a line to the counter time series.  The drift was determined to be $<9\times10^{-12}$ per day for the acetylene-referenced laser  and $<4\times10^{-11}$  for the hydrogen cyanide-referenced laser.  This corresponds to equivalent RV drifts of $<0.27$ cm s$^{-1}$ and $<1.2$ cm s$^{-1}$ per day for the two references. Finally, we attempted to place a bound on the repeatability of the 1542 nm reference laser during re-locking and power cycling.  Although only evaluated for a limited number of power cycles and re-locks, in all cases, we found that the laser frequency returned to its predetermined value within $<$100 kHz, or equivalently, with a RV imprecision of $<15$ cm s$^{-1}$. 

While these callibrations are sufficient for the few-day observations reported below, confidence in the longer term stability of the molecularly-referenced CW lasers would be required for observations that could extend over many years.  Properly addressing the potential frequency drifts on such a multi-year time scale would require a more thorough investigation of systematic frequency effects due to a variety of physical and operational parameters (e.g. laser power, pressure, temperature, and electronic offsets).  Alternatively, narrower absorption features, as available in nonlinear saturation spectroscopy, could provide improved performance.  For example, laboratory experiments have shown fractional frequency instability at the level of $10^{-12}$ and reproducibility of $1.5\times10^{-11}$ for lasers locked to a Doppler-free transition in acetylene \citep{Edwards2004}.

\subsection{Operational Considerations}
Operational robustness is as critical as performance.  As noted before, all components in this system are telecommunications off-the-shelf components. Pictures of the key components are shown in the left column of figure \ref{figure2}.  Polarization maintaining fiber based optical components are also used to eliminate polarization drift. Moreover, no complicated control-locking loops are involved in the system and no temperature control is required. Comb line spacing is set by a commercial, microwave oscillator that can be synchronized to a Rb clock for long-term stability. As a result, the comb is able to maintain its frequency, bandwidth and intensity without the need to adjust any parameters. In testing at IRTF, the intensity of individual comb teeth has been measured to deviate less than 2dB during a 5 day span, including multiple power-off and on cycling of the optical continuum generation system (see  figure \ref{figure4a}c). The repeatability of the comb spectrum makes possible application of spectral flattening techniques for post processing of the comb spectrum.   Several techniques, including a fiber Bragg grating or application of a commercial waveshaper, can be used to flatten the comb spectrum over a 100-200nm span, Flattening over wider span and larger intensity variation is possible by cascading these devices. 

\section{Telescope Demonstration and On Sky Results}

\subsection{IRTF Demonstration}
To demonstrate that the laser comb is portable, easy to use as a wavelength calibration standard, and stable over days, we shipped the laser comb to the NASA Infrared Telescope Facility (IRTF).   IRTF is a three-meter diameter infrared-optimized telescope located at the summit of Mauna Kea, Hawaii. The observatory houses a number of facility near-infrared spectroscopic and imaging capabilities and visitor instruments.  We have modified the CSHELL spectrograph, one of the facility instruments, to permit the addition of a fiber acquisition unit (FAU) for the injection of starlight and laser frequency comb light into a fiber array and focusing on the spectrograph entrance slit.

\subsubsection{Experimental Setup}
CSHELL is a cryogenic, near-infrared traditional slit-fed spectrograph, with a resolution of $R\sim\lambda/\Delta\lambda=46,000$ \citep{Greene1993, Tokunaga1990}.   CSHELL is mounted at the Cassegrain focus of IRTF.  Starlight is directed from the secondary mirror at a focal ratio of $f/37.5$ to a chromatic lens that brings the beam to a focus at f/15 at the entrance slit of CSHELL. CSHELL images a single $\sim$ 5 nm wide order spectrum on a 256$\times$256 InSb detector.  The observed order is selectable from a wavelength range of 1-5 $\mu$m.

In order for CSHELL to accommodate the injection of the laser frequency comb light via a single mode fiber, we made use of the FAU deployed on CSHELL in 2012 and described in detail in \citep{Plavchan2013a}.  We briefly summarize the experimental setup of the FAU herein.  Before the f/37.5 beam of star-light from the secondary mirror reaches the CSHELL entrance slit, the starlight is passed through an isotopic methane absorption gas cell to introduce a common optical path wavelength reference \citep{Plavchan2013b}.  A pickoff mirror is next inserted into the beam to re-direct the near-infrared starlight to a fiber via a fiber-coupling lens.  A dichroic window re-directs the visible light to a guide camera to maintain the position of the star on the entrance of the fiber tip.  For the starlight, we made use of a specialized non-circular core multi-mode fiber (MMF), with a 50$\times$100 micron rectangular core.   These fibers ``scramble'' the near-field spatial modes of the fiber so that the spectrograph is evenly illuminated by the output from the fiber, regardless of the alignment, focus, or weather conditions of the starlight impinging upon the input to the fiber.  We additionally made use of a dual-frequency agitator motor to vibrate most of the 10-meter length of the fiber to provide additional mode mixing, distributing the starlight evenly between all modes.  Finally, a lens and a second pickoff mirror are used to relay the output of the starlight from the fiber output back to the spectrograph entrance slit.  The entire optical path can be seen in Figure 3 in \citep{Plavchan2013b}.

We modified the FAU described in \citep{Plavchan2013b} to add a single mode fiber (SMF) carrying the laser comb light next to the non-circular core fiber carrying the starlight.  This was accomplished by replacing the output single fiber SMA  fiber chuck with a custom 3-D printed V-groove array ferrule. This allowed us to send the light from both the star and frequency comb to the entrance slit of the CSHELL spectrograph when rotated in the same orientation as the slit.

To match the power output of the laser comb light to the star-light, and to avoid damage to the InSb detector on CSHELL, a series of commercial, fiber-pigtailed 0-30 dB  variable attenuators were installed.

Finally, the laser comb and associated electronics rack were set up in the room temperature stabilized ($\sim\pm$ 5 $^{\circ}$C) control room.  A 50-meter length of single mode fiber was run from the control room to the telescope dome floor, and along the telescope mount to the CSHELL spectrograph to connect to the V-groove array and FAU.  The unpacking, setup and integration of the comb fiber with CSHELL was straightforward, and required only 2 days working at an oxygen-deprived elevation of 14,000 feet in preparation for the observing run.

\subsubsection{Results with CSHELL on IRTF}
Three partial nights of CSHELL telescope time in September 2014 were used for this first on-sky demonstration of the laser comb. Unfortunately, the observing run was plagued by poor weather conditions, with 5-10 magnitudes of extinction due to clouds.  Consequently, we observed the bright M2 II-III star $\beta$ Peg (H=-2.1 mag), which is a pulsating variable star (P=43.3 days).  Typical exposure times were 150 seconds, and multiple exposures were obtained in sequence,

The star was primarily observed at 1.67 $\mu$m, with and without the isotopic methane gas cell to provide a wavelength calibration comparison for the laser comb.  Other wavelengths were also observed to demonstrate that the spectral grasp of the comb is much larger than the spectral grasp of the spectrograph itself.

Given the obtained SNR (signal-to-noise ratio) on $\beta$ Peg from the high extinction due to clouds, and the limited spectral grasp, we anticipate a photon-noise limited radial velocity precision of ~15 m/s by co-adding all of the observations taken within a single night.  This will be sufficient to demonstrate that the comb is as stable a wavelength calibration reference as the absorption gas cell.  However, the SNR is inadequate to demonstrate that the comb is more stable than the gas cell, as has been demonstrated in the lab (Section 2).

\begin{figure}[h]
\centering
\includegraphics[width=0.9\linewidth,natwidth=906,natheight=308]{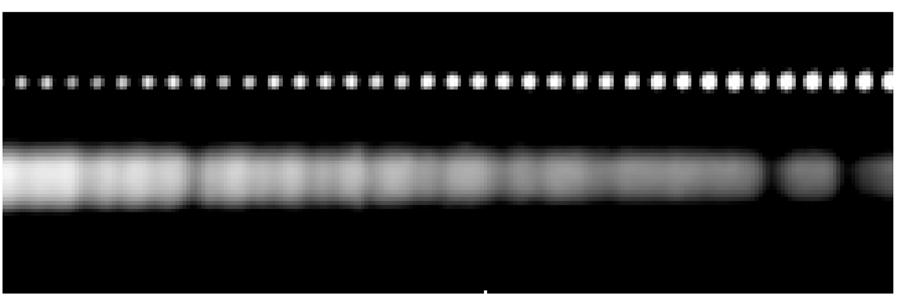}
\includegraphics[width= 0.6\linewidth,natwidth=2000,natheight=1200]{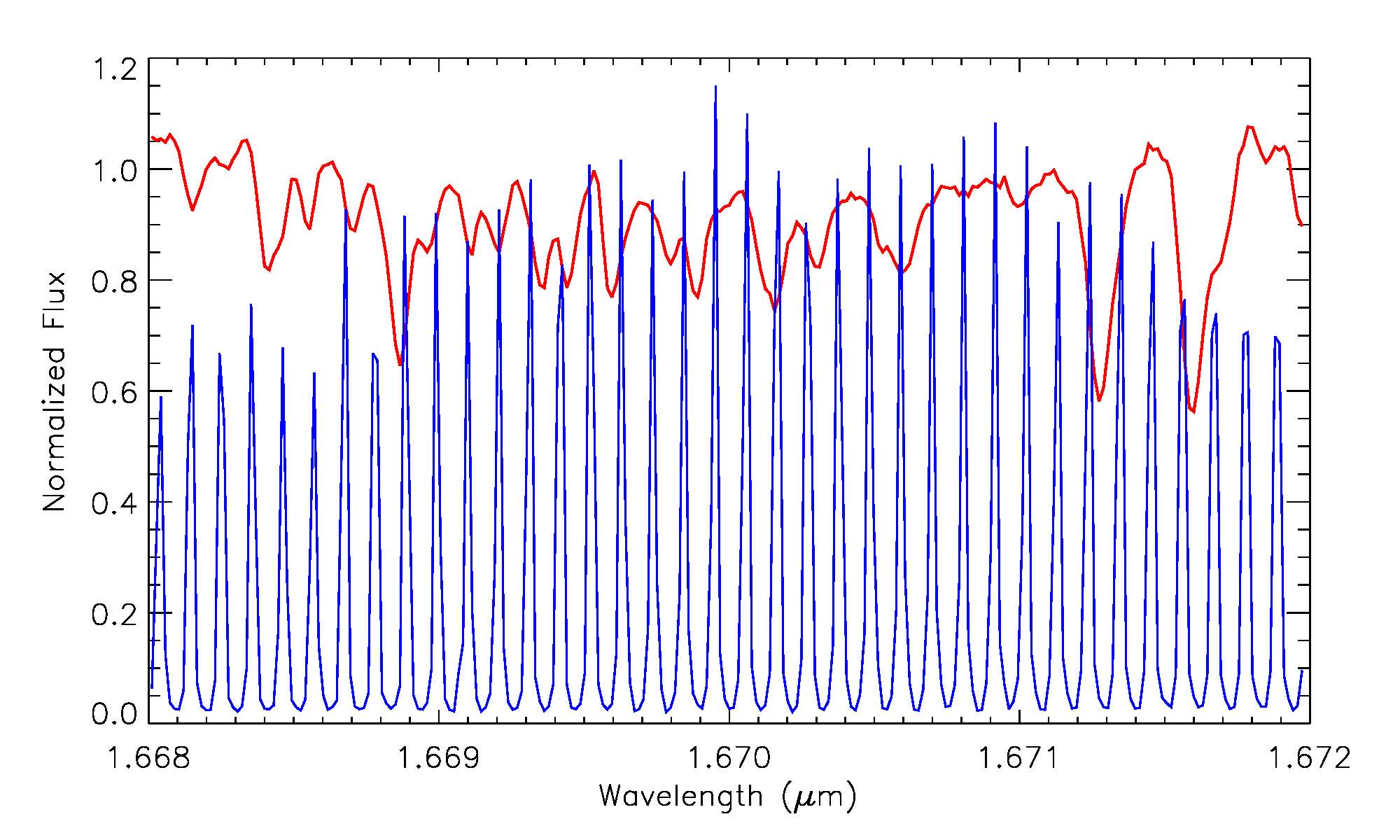}
\caption{Experimental results at IRTF. Upper: An image of the echelle spectrum from CSHELL on IRTF showing a 4 nm portion of spectrum around 1670 nm. The top row of dots are the laser comb lines while the broad spectrum at the bottom is from the bright M2 II-III giant star $\beta$ Peg seen through dense cloud cover. Lower: Spectra extracted from the above image. The solid red curve denotes the average of 11 individual spectra of $\beta$  Peg (without the gas cell) obtained with CSHELL on the IRTF. The regular sine-wave like blue lines show the  spectrum from the laser comb obtained simultaneously with the stellar spectrum.  The vertical axis is normalized flux units.}
 \label{betaPeg2dim}\label{betaPegspectrum}
\end{figure}

One critical aspect of demonstrating the usability of the comb for astrophysical spectrographs is the comb line spacing.  As seen in Figures \ref{betaPeg2dim} and insets of fig \ref{figure4a} (b), the spectra clearly demonstrate that the individual comb lines are resolved with the CSHELL spectrograph without the need for additional line filtering, e.g. \citet{Osterman2011}.  Thus this comb operates at a frequency that is natively well-suited for astronomical applications with significantly less hardware complexity compared to ``traditional'' laser frequency combs. Another critical aspect of wavelength stability in astrophysical applications is the stability of the illumination of the spectrograph. Single mode fiber (SMF is ideal for spectrograph illumination, as only a single Gaussian mode is propagated through the fiber.  As long as the SMF is mechanically stable, the spectrograph illumination will be as well.  For comparison, the MMF shows less illumination stability, even with the specialized non-circular core rectangular fiber.  This is likely due to the limited number of spatial modes in the fiber at these near-infrared wavelengths.  However, with the fiber agitation turned on, the illumination of the MMF approaches the stability of the SMF.

\subsection{Keck Telescope Demonstration}

\begin{figure}[h]
\centering
\includegraphics[width=0.9\linewidth,natwidth=577,natheight=454]{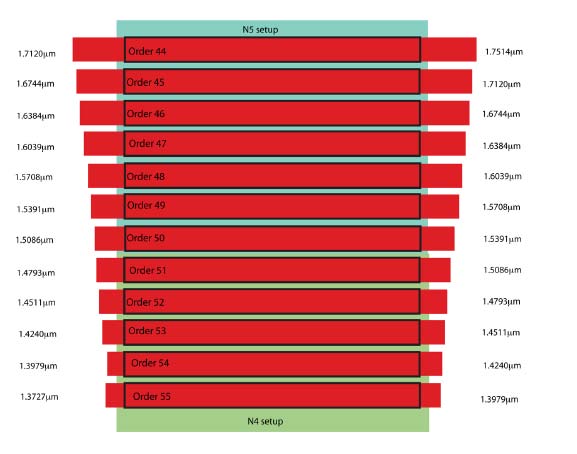}

\caption{A schematic layout of the NIRSPEC echelle as set up for wavelengths in the H band. Approximately 10 orders are observable at one time spanning most of the H band at high resolution. In the configuration without the EDFA amplifier, the  comb generated with acetelyne and hydrogen cyanide lasers illuminated two 4 nm sections of a single order. In full operation, the comb would fill over 2 orders around 1550 nm.}
 \label{NIRSPEC-Layout}
\end{figure}

We were able to use daytime access to the NIRSPEC spectrometer on the Keck-II telescope \citep{McLean1998} to demonstrate both  acetelyne- and HCN-based laser combs. NIRSPEC is a cross-dispersed echelle capable of covering a large fraction of the entire H-band in a single setting with a spectral resolution of R$\sim$25,000 (Figure \ref{NIRSPEC-Layout}). We injected the comb signal using a fiber feed into the integrating sphere at the input to the NIRSPEC calibration subsystem. While this arrangement did not allow for simultaneous stellar and comb observations, we were able to measure the comb lines simultaneously with the arc lamps normally used for wavelength calibration and to make hour-long tests of the stability of the NIRSPEC instrument at the sub-nm level. Unfortunately the EDFA amplifier was not availble for these tests so that we were restricted to combs spanning only 4 nm instead of the broader combs described above.

\begin{figure}[h]
\centering
\includegraphics[width=0.9\linewidth,natwidth=800,natheight=600]{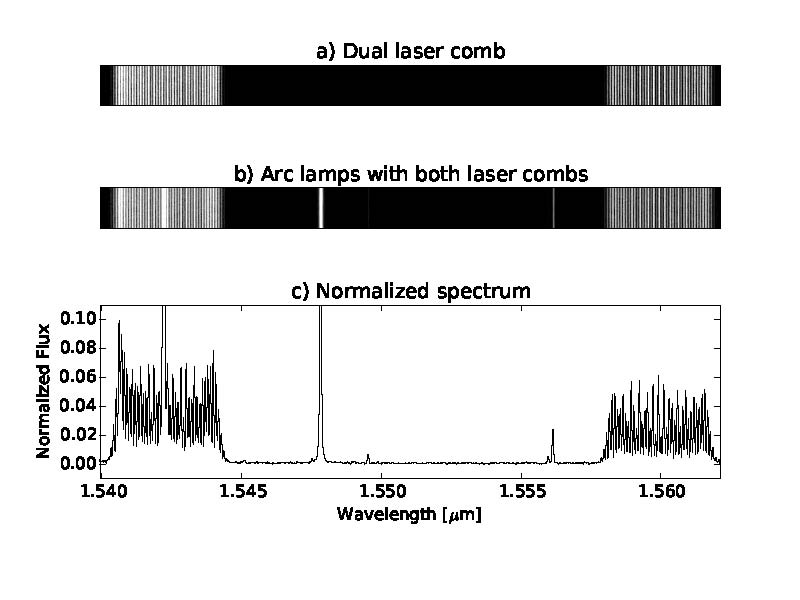}

\caption{Data from testing at Keck II. (Upper) Reduced NIRSPEC image from echelle order 49 displaying both acetylene (left) and HCN (right) stabilized laser combs. (Middle) Reduced NIRSPEC image from echelle order 49 with Xe, Ne, Ar, and Kr arc lamps and both laser combs. (Lower) Normalized spectrum summed over the 10 central pixel rows from the middle image.}
 \label{NIRSPEC-comb}
\end{figure}

An echelleogram (upper panel in Figure \ref{NIRSPEC-comb}) shows the result of illuminating the integrating sphere with both laser combs simultaneously. The two bright areas in the image are approximately 4 nm wide and centered on 1542 and 1560 nm for  acetylene and hydrogen cyanide stabilized combs, respectively. The middle panel shows the combination of the laser combs along with the  lines from the arc lamps (Kr, Ne, Ar and Xe) used for wavelength calibration. The echelleograms were reduced in standard fashion, correcting for dark current and flat field variations. Spectral extractions  (lower panel in Figure \ref{NIRSPEC-comb}) show that  individual comb lines are well resolved at NIRSPEC's resolution and spaced approximately 4 pixels apart (0.1 nm), consistent with the higher resolution IRTF observations described above.

Eventual operation with a laser comb covering over $>$ 90 nm of the H band would allow much higher radial velocity precision than is presently possible using, for example, atmosperic OH lines, as a wavelength standard. NIRSPEC's ultimate RV precision will depend on many factors, including the brightness of the star, NIRSPEC's spectral resolution (presently 25,000 but increasing to 50,000 after a proposed upgrade) and the ability to stabilize the input stellar light against pointing drifts and line profile variations. We anticipate that in an exposure of 900 sec NIRSPEC should be able to achieve an RV precision $\sim$ 1 m s$^{-1}$ for a large number of nearby M stars brighter than H$\sim$10 mag. A stable wavelength reference, observed simultaneously with the stellar spectrum, is critical to achieving this precision.

\section{Future Prospects}

Many challenges remain to achieving the high precision RV capability needed for the study of exoplanets orbiting late M dwarfs or jitter-prone hotter G and K spectral types. Achieving adequate signal to noise on relatively faint stars requires a large spectral grasp on a high-resolution spectrometer on a large aperture telescope. Injecting both the laser comb and starlight into the spectrograph with a highly stable line spread function demands carefully designed  interfaces between the comb light and star light at the entrance to the spectrograph. Extracting the data from the spectrometer requires careful attention to flat-fielding and other detector features. Finally, reducing the extracted spectra to produce RV measurements at the required level of precision requires sophisticated modeling of complex stellar atmospheres and telluric atmospheric absorption. The research described here addresses only one of these steps, namely the generation of a highly stable wavelength standard in the NIR suitable for sub m s$^{-1}$ RV detection.

\section*{Acknowledgments}
Three IRTF nights were donated in September 2014 to integrate and test the laser comb with CSHELL.  One of these nights came from IRTF engineering time and the other two came from Peter Plavchan's CSHELL program to observe nearby M dwarfs with the absorption gas cell to obtain precise radial velocities. We are grateful to the leadership of the IRTF, Director Alan Tokunaga and Deputy Director John Rayner, as well as to the daytime and night time staff at the summit for their support. We further thank Jeremy Colson at Wavelength References for his assistance with the molecular-stabilized lasers.  

On-sky observations were obtained at the Infrared Telescope Facility, which is operated by the University of Hawaii under Cooperative Agreement no. NNX-08AE38A with the National Aeronautics and Space Administration, Science Mission Directorate, Planetary Astronomy Program.

Daytime operations at the Keck-II telescope were carried out with the assistance of Sean Adkins and Steve Milner. We greatfully acknowledge the support of the entire Keck summit team in making these tests possible.

The authors wish to recognize and acknowledge the very significant cultural role and reverence that the summit of Mauna Kea has always had within the indigenous Hawaiian community.  We are most fortunate to have the opportunity to conduct observations from this mountain. The data presented herein were obtained at the W.M. Keck Observatory, which is operated as a scientific partnership among the California Institute of Technology, the University of California and the National Aeronautics and Space Administration. The Observatory was made possible by the generous financial support of the W.M. Keck Foundation.

SD and GY acknowledge support from NIST and the NSF grant AST-1310875.

This research was carried out at the Jet Propulsion Laboratory and the California Institute of Technology under a contract with the National Aeronautics and Space Administration and funded through the President’s and Director’s Fund Program.  Copyright 2014 California Institute of Technology. All rights reserved.





\clearpage

\clearpage

\clearpage

\clearpage

\clearpage

\end{document}